\documentclass{iopart}

\begin{document}

% J. Phys. CM identifier
\jl{3}

\letter{Quadratic short-range order corrections to the mean-field free energy
%\footnote[1]{Preprint cond-mat/NNNNNN in the LANL e-print archive.}
}

\author{Igor Tsatskis\footnote[1]{E-mail: it10001@cus.cam.ac.uk; 
former name: I V Masanskii.}}

\address{Department of Earth Sciences, University of Cambridge, Downing Street,\\
Cambridge CB2 3EQ, United Kingdom}

\begin{abstract}
A method for calculating the short-range order part of the free energy of
order-disorder systems is proposed. The method is based on the apllication
of the cumulant expansion to the exact configurational entropy. Second-order
correlation corrections to the mean-field approximation for the free energy
are calculated for arbitrary thermodynamic phase and type of interactions.
The resulting quadratic approximation for the correlation entropy leads to
substantially better values of transition temperatures for the
nearest-neighbour cubic Ising ferromagnets.
\end{abstract}

\submitted

\nosections
In a recent paper~\cite{Tsatskis-EPL} we developed a method of deriving very 
simple polynomial approximations for the variational configurational entropy
in which the well-known and reliable cluster variation method (CVM) of
Kikuchi~\cite{Kikuchi} was used as a starting point. It was observed that in
the absolute majority of the CVM approximations the number of variational
parameters in the entropy is substantially greater than in the internal
energy. As a result, the minimization of the free energy with respect to
those variables which are not present in the internal energy reduces to the
maximization of the entropy. The corresponding equations simply relate
different variational parameters, since the configurational entropy contains
no information about the character and strength of interactions in the
system. Most of the variables in the CVM free energy are therefore redundant
and can in principle be eliminated by solving these equations. However, the
equations are non-linear and apparently cannot be solved exactly; even if
this was possible, the resulting expression for the entropy would be far too
complicated. To overcome this difficulty, it was proposed to expand all the
redundant parameters in powers of cumulants of those pair and higher-order
averages which enter the expression for the internal energy; such cumulants
are usually sufficiently small. In this case the initial equations reduce to
those for the corresponding expansion coefficients, and the latter can be
easily solved. The final approximation for the entropy is a polynomial of
some order in the cumulants of only those averages which are present in the
internal energy. The proposed method can be regarded as a simplification of
the CVM.

The described approach, however, combines two very different approximations,
since its aim is to find an approximate analytical solution of the already
approximate CVM equations (which are usually treated only numerically). In
the present letter we show that it is possible to apply the same idea of the
cumulant expansion directly to the exact configurational entropy, thus
avoiding using the CVM for constructing the initial approximation. The
general formalism is presented and the first non-zero (quadratic) 
corrections due to short-range order (SRO) to the mean-field approximation 
(MFA) for the free energy are calculated. Considerably more complicated 
calculation of higher-order terms is left for a discussion 
elsewhere~\cite{Tsatskis-to-be-published}. It is shown that the obtained 
quadratic approximation for the correlation entropy significantly improves, 
in comparison with the MFA, calculated transition temperatures for the
nearest-neighbour (NN) cubic Ising ferromagnets.

The formally exact expression for the configurational entropy has the form
\begin{equation}
S = - k_B \sum_{i_1 \ldots i_N} P_{1 \ldots N}^{i_1 \ldots i_N} 
\ln P_{1 \ldots N}^{i_1 \ldots i_N} . \label{eq1}
\end{equation}
Here $k_B$ is the Boltzmann constant, the $N$-site probability
\begin{equation}
P_{1 \ldots N}^{i_1 \ldots i_N} = \left\langle p_1^{i_1} \ldots
p_N^{i_N} \right\rangle \label{eq2}
\end{equation}
is the statistical average of the product of $N$ occupation numbers, 
and $N$ is the number of lattice sites in the system. The occupation 
number $p_n^i$ for a particular spin orientation (up or down) is equal 
to $1$ or $0$ depending on whether the spin at the lattice site $n$ 
has this orientation or not. It is convenient for our purposes to 
define the orientation index $i$ as taking on values $\pm 1$. 
Equation~(\ref{eq1}) is a slightly more detailed version of the 
familiar result~\cite{Landau-Lifshitz} 
\begin{equation}
S = - k_B \mbox{Tr} \rho \ln \rho \label{eq3}
\end{equation}
where $\rho$ is the density matrix and Tr denotes the trace of a matrix.

The cumulant average~\cite{Kubo} of a product of $l$ variables is the 
$l$th-order residue when all combinations of cumulant averages of lower order
are subtracted from the actual average of the product. The cumulant averages
of the occupation numbers are thus defined by the relations
\begin{equation}
\eqalign{
\fl \left\langle p_{n_1}^{j_1} \right\rangle = 
\left\langle p_{n_1}^{j_1} \right\rangle_c \nonumber \\
\fl \left\langle p_{n_1}^{j_1} p_{n_2}^{j_2} \right\rangle = 
\left\langle p_{n_1}^{j_1} \right\rangle_c 
\left\langle p_{n_2}^{j_2} \right\rangle_c
+\left\langle p_{n_1}^{j_1} p_{n_2}^{j_2} \right\rangle_c \nonumber \\
\fl \left\langle p_{n_1}^{j_1} p_{n_2}^{j_2} p_{n_3}^{j_3} \right\rangle =
\left\langle p_{n_1}^{j_1} \right\rangle_c
\left\langle p_{n_2}^{j_2} \right\rangle_c
\left\langle p_{n_3}^{j_3} \right\rangle_c
+\left\langle p_{n_1}^{j_1} p_{n_2}^{j_2} \right\rangle_c 
\left\langle p_{n_3}^{j_3} \right\rangle_c 
+\left\langle p_{n_1}^{j_1} p_{n_3}^{j_3} \right\rangle_c
\left\langle p_{n_2}^{j_2} \right\rangle_c \nonumber \\
+\left\langle p_{n_2}^{j_2} p_{n_3}^{j_3} \right\rangle_c
\left\langle p_{n_1}^{j_1} \right\rangle_c 
+\left\langle p_{n_1}^{j_1} p_{n_2}^{j_2} p_{n_3}^{j_3} 
\right\rangle_c \label{eq4}
}
\end{equation}
etc., where the subscript $c$ denotes a cumulant average. Correspondingly,
the $N$-site probability~(\ref{eq2}) is written as 
\begin{eqnarray}
\fl P_{1 \ldots N}^{i_1 \ldots i_N} = 
\left\langle p_1^{i_1} \right\rangle_c \ldots 
\left\langle p_N^{i_N} \right\rangle _c 
+\Big[ \left\langle p_1^{i_1} p_2^{i_2} \right\rangle_c
\left\langle p_3^{i_3} \right\rangle_c \ldots 
\left\langle p_N^{i_N} \right\rangle_c + \ldots \nonumber \\
+\left\langle p_1^{i_1} \right\rangle_c \ldots 
\left\langle p_{N-2}^{i_{N-2}} \right\rangle_c 
\left\langle p_{N-1}^{i_{N-1}} p_N^{i_N} \right\rangle_c \Big] \nonumber \\
+\Big[ \left\langle p_1^{i_1} p_2^{i_2} p_3^{i_3} \right\rangle_c
\left\langle p_4^{i_4} \right\rangle_c \ldots 
\left\langle p_N^{i_N} \right\rangle_c + \ldots \nonumber \\
+\left\langle p_1^{i_1} \right\rangle_c \ldots 
\left\langle p_{N-3}^{i_{N-3}} \right\rangle_c 
\left\langle p_{N-2}^{i_{N-2}} p_{N-1}^{i_{N-1}} p_N^{i_N} 
\right\rangle_c \Big] + \ldots \nonumber \\
+\Big[ \left\langle p_1^{i_1} p_2^{i_2} \right\rangle_c 
\left\langle p_3^{i_3} p_4^{i_4} \right\rangle_c
\left\langle p_5^{i_5} \right\rangle_c \ldots 
\left\langle p_N^{i_N} \right\rangle_c + \ldots \nonumber \\
+\left\langle p_1^{i_1} \right\rangle_c \ldots 
\left\langle p_{N-4}^{i_{N-4}} \right\rangle_c 
\left\langle p_{N-3}^{i_{N-3}} p_{N-2}^{i_{N-2}} \right\rangle_c
\left\langle p_{N-1}^{i_{N-1}} p_N^{i_N} 
\right\rangle_c \Big] + \ldots \nonumber \\
+\left\langle p_1^{i_1} \ldots p_N^{i_N} \right\rangle_c . \label{eq5}
\end{eqnarray}
Introducing notations 
\begin{eqnarray}
P_{n_1 \ldots n_q}^{j_1 \ldots j_q} = \left\langle p_{n_1}^{j_1} \ldots
p_{n_q}^{j_q} \right\rangle \qquad & C_{n_1 \ldots n_q}^{j_1 \ldots j_q}
= \left\langle p_{n_1}^{j_1} \ldots p_{n_q}^{j_q} \right\rangle_c
\label{eq6}
\end{eqnarray}
and noticing that $P_n^j=C_n^j$ as given by the first of equations~(\ref{eq4}),
we rewrite the rest of equations~(\ref{eq4}) and equation~(\ref{eq5}) as follows, 
\begin{equation}
\eqalign{
\fl P_{n_1n_2}^{j_1j_2} = P_{n_1}^{j_1}P_{n_2}^{j_2}
+C_{n_1n_2}^{j_1j_2} \nonumber \\
\fl P_{n_1n_2n_3}^{j_1j_2j_3} = P_{n_1}^{j_1}P_{n_2}^{j_2}P_{n_3}^{j_3}
+C_{n_1n_2}^{j_1j_2}P_{n_3}^{j_3}+C_{n_1n_3}^{j_1j_3}P_{n_2}^{j_2} 
+C_{n_2n_3}^{j_2j_3}P_{n_1}^{j_1}+C_{n_1n_2n_3}^{j_1j_2j_3} \nonumber \\
\fl \ldots \nonumber \\
\fl P_{1 \ldots N}^{i_1 \ldots i_N} = P_1^{i_1} \ldots P_N^{i_N}+\left(
C_{12}^{i_1i_2}P_3^{i_3} \ldots P_N^{i_N} + \ldots + P_1^{i_1} \ldots 
P_{N-2}^{i_{N-2}}C_{N-1,N}^{i_{N-1}i_N}\right) \nonumber \\
+\left( C_{123}^{i_1i_2i_3}P_4^{i_4} \ldots P_N^{i_N} + \ldots 
+P_1^{i_1} \ldots P_{N-3}^{i_{N-3}}
C_{N-2,N-1,N}^{i_{N-2}i_{N-1}i_N} \right) + \ldots \nonumber \\
+\left( C_{12}^{i_1i_2}C_{34}^{i_3i_4}P_5^{i_5} \ldots P_N^{i_N} + \ldots 
+P_1^{i_1} \ldots P_{N-4}^{i_{N-4}}C_{N-3,N-2}^{i_{N-3}i_{N-2}}
C_{N-1,N}^{i_{N-1}i_N}\right) \nonumber \\
+ \ldots +C_{1 \ldots N}^{i_1 \ldots i_N} . \label{eq7}
}
\end{equation}
Finally, defining new quantities 
\begin{equation}
Q_{n_1 \ldots n_q}^{j_1 \ldots j_q}=\frac{C_{n_1 \ldots n_q}^{j_1 \ldots j_q}}{
P_{n_1}^{j_1} \ldots P_{n_q}^{j_q}}  \label{eq8}
\end{equation}
we represent equations~(\ref{eq7}) in the form 
\begin{equation}
\eqalign{
P_{n_1n_2}^{j_1j_2} = P_{n_1}^{j_1}P_{n_2}^{j_2}\left(
1+Q_{n_1n_2}^{j_1j_2}\right)  \nonumber \\
P_{n_1n_2n_3}^{j_1j_2j_3} = P_{n_1}^{j_1}P_{n_2}^{j_2}P_{n_3}^{j_3}\left(
1+Q_{n_1n_2}^{j_1j_2}+Q_{n_1n_3}^{j_1j_3}+Q_{n_2n_3}^{j_2j_3}
+Q_{n_1n_2n_3}^{j_1j_2j_3}\right) \nonumber \\
\ldots \nonumber \\
P_{1 \ldots N}^{i_1 \ldots i_N} = P_1^{i_1} \ldots P_N^{i_N}\left(
1+X_{1 \ldots N}^{i_1 \ldots i_N}\right) \label{eq9}
}
\end{equation}
where in the last equation
\begin{equation}
\fl X_{1 \ldots N}^{i_1 \ldots i_N} = \sum_{q\geq 2}\sum_{\left\langle 
n_1 \ldots n_q\right\rangle }Q_{n_1 \ldots n_q}^{j_1 \ldots j_q} 
+\sum_{q\geq 2}\sum_{p\geq 2}\sum_{\left\langle n_1 \ldots n_{q+p}
\right\rangle}Q_{n_1 \ldots n_q}^{j_1 \ldots j_q}Q_{n_{q+1} \ldots 
n_{q+p}}^{j_{q+1} \ldots j_{q+p}}+O\left( C^3\right) . \label{eq10}
\end{equation}
Here notation $\left\langle n_1 \ldots n_q\right\rangle $ corresponds to
summation over clusters consisting of lattice sites $n_1, \, \ldots, \, n_q$,
and the last term denotes all contributions containing products of three or
more cumulants $C_{n_1 \ldots n_q}^{j_1 \ldots j_q}$. Insertion of the result
for the $N$-site probability (the last of equations~(\ref{eq9})) into 
equation~(\ref{eq1}) leads to the natural separation of the long-range order 
(MFA) and SRO (correlation) contributions to the configurational entropy, 
\numparts
\begin{eqnarray}
S = S_0+S_1 \label{eq11} \\
S_0 = -k_B\sum_{in}P_n^i\ln P_n^i \label{eq12} \\
S_1 = -k_B\sum_{i_1 \ldots i_N}P_1^{i_1} \ldots P_N^{i_N} \left(
1+X_{1 \ldots N}^{i_1 \ldots i_N}\right) \ln \left( 1+X_{1 \ldots N}^{i_1 \ldots
i_N}\right) . \label{eq13}
\end{eqnarray}
\endnumparts
We now expand the correlation entropy $S_1$ in powers of $X_{1 \ldots
N}^{i_1 \ldots i_N}$. It can be shown that the linear term vanishes \cite
{Tsatskis-to-be-published}; this result is obvious, since the entropy is
maximal in the absence of correlations. Substitution of equation~(\ref{eq10})
into equation~(\ref{eq13}) then gives 
\begin{eqnarray}
\fl S_1 = -\frac 12k_B\sum_{i_1 \ldots i_N}P_1^{i_1} \ldots P_N^{i_N}
\sum_{q\geq 2}\sum_{p\geq 2} 
\sum_{\left\langle n_1 \ldots n_q\right\rangle}
\sum_{\left\langle m_1 \ldots m_p\right\rangle }Q_{
n_1 \ldots n_q}^{j_1 \ldots j_q}Q_{m_1 \ldots m_p}^{k_1 \ldots k_p} 
+O\left( C^3\right) . \label{eq14}
\end{eqnarray}
The next step is to use the relation 
\begin{equation}
C_{n_1 \ldots n_q}^{j_1 \ldots j_q}=2^{-q}\ j_1 \ldots j_q\ C_{n_1 \ldots n_q}
\label{eq15}
\end{equation}
where $C_{n_1 \ldots n_q}$ is the cumulant average of the product of the
corresponding spin variables $s_n$ which are linearly related to the
occupation numbers and acquire values $+1$ for spin up and $-1$ for spin
down, 
\begin{equation}
C_{n_1 \ldots n_q}=\left\langle s_{n_1} \ldots s_{n_q}\right\rangle _c .
\label{eq16}
\end{equation}
The proof of equation~(\ref{eq15}) is not given here to save space \cite
{Tsatskis-to-be-published}, but its validity can be directly checked, at
least for low-order cumulants. Inserting equation~(\ref{eq15}) into 
equation~(\ref{eq8}) and then the result into equation~(\ref{eq14}), we obtain 
\begin{eqnarray}
\fl S_1 = -k_B\sum_{q\geq 2}\sum_{p\geq 2}2^{-(q+p+1)}\sum_{\left\langle 
n_1 \ldots n_q\right\rangle }\sum_{\left\langle m_1 \ldots m_p\right\rangle } 
\left(\sum_{i_1 \ldots i_N}\frac{j_1 \ldots j_q\,k_1 \ldots k_p\,P_1^{i_1}
\ldots P_N^{i_N}}{P_{n_1}^{j_1} \ldots P_{n_q}^{j_q}\,P_{m_1}^{k_1}
\ldots P_{m_p}^{k_p}} \right) \nonumber \\
\times C_{n_1 \ldots n_q}C_{m_1 \ldots m_p} +O\left( C^3\right) . \label{eq17}
\end{eqnarray}
The expression in the round brackets in equation~(\ref{eq17}) factorizes into the
product of sums over orientations of individual spins. The result of each
such summation depends on whether the corresponding lattice site belongs to
both sets $n_1, \, \ldots, \, n_q$ and $m_1, \, \ldots, \, m_p$, to only one
of them, or to neither, and is 
\numparts 
\begin{eqnarray} 
\sum_jj^2(P_n^j)^{-1} = (+1)^2(P_n^{+})^{-1}+(-1)^2(P_n^{-})^{-1}=\left(
P_n^{+}P_n^{-}\right) ^{-1} \label{eq18} \\
\sum_jj = (+1)+(-1)=0 \label{eq19} \\
\sum_jP_n^j = P_n^{+}+P_n^{-}=1 \label{eq20}
\end{eqnarray}
\endnumparts
for these three cases, respectively (in the second case the probabilities in
the numerator and denominator cancel each other). Equation~(\ref{eq19}) shows
that the only surviving terms in equation~(\ref{eq17}) are those in which the
sets $n_1, \, \ldots, \, n_q$ and $m_1, \, \ldots, \, m_p$ represent the same
lattice cluster. Substituting equations~(\ref{eq18})-(\ref{eq20}) into 
equation~(\ref{eq17}) and neglecting products of three or more cumulants 
$C_{n_1 \ldots n_q}$, we finally obtain the expression for the correlation 
entropy in the second-order approximation, 
\begin{equation}
S_1=-k_B\sum_{q\geq 2}2^{-(2q+1)}\sum_{\left\langle n_1 \ldots
n_q\right\rangle }\left( P_{n_1}^{+}P_{n_1}^{-} \ldots
P_{n_q}^{+}P_{n_q}^{-}\right) ^{-1}C_{n_1 \ldots n_q}^2 . \label{eq21}
\end{equation}
In this order the correlation entropy is thus a linear combination of
squares of the cumulants, i.e., contains no cross-terms. The minimization 
of the free energy then leads to vanishing of all those cumulants which 
are not present in the internal energy. 

Equation~(\ref{eq21}) for the correlation entropy has been obtained
without any assumptions about the nature of a thermodynamic phase or
interactions in the system. We now list several frequently encountered
situations in which the result for the free energy has particularly simple
form.

{\em (i)} In the absence of sublattices, when all lattice sites are
equivalent (such as in the case of disordered or ferromagnetically ordered
material), the point probabilities $P_n^i$ do not depend on the site index $n
$, 
\begin{equation}
P_n^{+}=\frac{1+m}2 \qquad P_n^{-}=\frac{1-m}2 \label{eq22}
\end{equation}
where $m=\left\langle s_n\right\rangle $ is the site magnetization. 
Equations~(\ref{eq12}) and (\ref{eq21}) therefore become 
\numparts
\begin{eqnarray}
S_0 = -Nk_B\left( \frac{1+m}2\ln \frac{1+m}2+\frac{1-m}2\ln \frac{1-m}2
\right) \label{eq23} \\
S_1 = -\frac 12k_B\sum_{q\geq 2}\!\left( 1-m^2\right)
^{-q}\sum_{\left\langle n_1 \ldots n_q\right\rangle }C_{n_1 \ldots n_q}^2 .
\label{eq24}
\end{eqnarray}
\endnumparts

{\em (ii)} If pairwise interactions are assumed in the case {\em (i)}, then
the only terms necessary to retain in equation~(\ref{eq24}) are those with $q=2$.
All other, higher-order cumulants vanish at equilibrium, since the internal
energy is independent of corresponding averages. As a result, 
\begin{equation}
S_1=-\frac 12 k_B \left( 1-m^2\right) ^{-2}\sum_{\left\langle
nm\right\rangle }C_{nm}^2 . \label{eq25}
\end{equation}
At this stage it is convenient to use pair SRO parameters $\alpha _{nm}$
instead of pair cumulants $C_{nm}$; the relation between them is \cite
{Tsatskis-Michigan} 
\begin{equation}
C_{nm}=\left( 1-m^2\right) \alpha _{nm} . \label{eq26}
\end{equation}
In terms of the SRO parameters equation~(\ref{eq25}) has the form 
\begin{equation}
S_1=-\frac 12k_B\sum_{\left\langle nm\right\rangle }\alpha _{nm}^2=-\frac 14
Nk_B\sum_lZ_l\alpha _l^2 \label{eq27}
\end{equation}
where in the second equation $Z_l$ and $\alpha _l$ are the coordination
number and SRO parameter for the coordination shell $l$, and the summation
is performed over all coordination shells.

{\em (iii)} When only the NN spins interact in the case {\em (ii)}, it is
sufficient to take into account in the second of equations~(\ref{eq27}) only the
term corresponding to the first coordination shell. The rest of the SRO
parameters vanish at equilibrium for the same reason as before. The
correlation entropy in this case is simply 
\begin{equation}
S_1=-\frac 14Nk_BZ\alpha ^2 . \label{eq28}
\end{equation}
Here we have dropped the subscripts in the NN coordination number and in the
NN SRO parameter. For the FCC lattice ($Z=12$) this result coincides with
that obtained in \cite{Tsatskis-EPL} within the tetrahedron CVM
approximation. This means that the tetrahedron version of the CVM reproduces
correctly the exact expansion for the configurational entropy up to at least
second order in $\alpha $.

Finally, to test the obtained approximation for the free energy, we consider
thermodynamics of the NN Ising ferromagnet, an example which corresponds to
the simplest case {\em (iii)}. The Hamiltonian of the Ising model is written
as 
\begin{equation}
H=-\sum_{\left\langle nm\right\rangle }J_{nm}s_ns_m-h\sum_ns_n .
\label{eq29}
\end{equation}
Here $J_{nm}$ is the exchange integral equal to $J>0$ for NN sites and to
zero otherwise, and $h$ is the magnetic field. The internal energy, being
the statistical average of the Hamiltonian, is therefore 
\begin{equation}
E=-\frac 12NZJ\left[ m^2+\left( 1-m^2\right) \alpha \right] -Nhm
\label{eq30}
\end{equation}
where the first two of equations~(\ref{eq4}) for the spin variables and 
equation~(\ref{eq26}) were used. Combining equations~(\ref{eq23}), 
(\ref{eq28}) and (\ref{eq30}), we get for the free energy $F=E-TS$ 
($T$ is the absolute temperature) 
\begin{eqnarray}
\fl F = N\left[ \left( -\frac 12ZJm^2-hm\right) -\frac 12ZJ
\left( 1-m^2\right) \alpha \right. \nonumber \\
\left. +k_BT\left( \frac{1+m}2\ln \frac{1+m}2+\frac{1-m}2\ln 
\frac{1-m}2\right) +\frac 14k_BTZ\alpha ^2\right] . \label{eq31}
\end{eqnarray}
The four terms in equation~(\ref{eq31}), left to right, are the MFA internal
energy, correlation internal energy, MFA entropy and correlation entropy,
respectively. The variational free energy is now a function of two
minimization parameters $m$ and $\alpha $, whose equilibrium values are
determined by the conditions 
\begin{equation}
\frac{\partial F}{\partial m}=0 \qquad \frac{\partial F}{\partial \alpha }=0 .
\label{eq32}
\end{equation}
They lead to the following equations for $m$ and $\alpha $, 
\begin{equation}
m=\tanh \left[ \frac{m(1-\alpha )+\mu }t\right] \qquad \alpha =\frac{\left(
1-m^2\right) }{Zt} \label{eq33}
\end{equation}
where $t=k_BT/ZJ$ and $\mu =h/ZJ$ are the dimensionless temperature and
magnetic field. Note that the first of equations~(\ref{eq33}) differs from its
well-known MFA counterpart by the factor $1-\alpha $. The temperature of the
second-order transition is the solution of the equation 
\begin{equation}
\left( \frac{\partial ^2F}{\partial m^2}\right) _{m=0}=0 . \label{eq34}
\end{equation}
Combining this equation with the second of equations~(\ref{eq33}) for $m=0$, we
obtain the equation for $t_c$, 
\begin{equation}
t_c=1-\left( Zt_c\right) ^{-1} .  \label{eq35}
\end{equation}
The $Z$-independent MFA result $t_c=1$ is obtained from this equation in the
limit $Z\rightarrow \infty $. Equation~(\ref{eq35}) shows that the derived SRO
corrections to the free energy lead to the decrease in $t_c$ of order $Z^{-1}
$, as expected \cite{Brout}. Results of the numerical solution of 
equation~(\ref{eq35}) for the three cubic lattices are given in table~\ref{t1}; 
they are compared with the best-known $t_c$ values \cite{Domb}. It is seen 
that the quadratic approximation for the correlation entropy improves the 
accuracy of the calculated $t_c$ from 20-30\% in the MFA to 5-10\%, and the 
accuracy increases with decreasing coordination number.

\Table{Normalized temperature $t_c=k_BT_c/ZJ$ of the second-order phase transition
for the NN Ising ferromagnet on the simple cubic (SC), body-centered cubic
(BCC), and face-centered cubic (FCC) lattices calculated using 
equation~(\ref{eq35}), i.e., within the quadratic approximation (\ref{eq28}) 
for the correlation entropy. The best-known $t_c$ values (which can be 
considered as essentially exact), the MFA results and the corresponding 
ratios are shown for comparison.\label{t1}}
\br
Lattice & $Z$ & $t_c$ & $t_c^{MFA}$ & $t_c^{exact}$ & $t_c/t_c^{exact}$ & 
$t_c^{MFA}/t_c^{exact}$ \\
\mr
SC & 6 & 0.789 & 1 & 0.752 & 1.049 & 1.330 \\ 
BCC & 8 & 0.854 & 1 & 0.794 & 1.075 & 1.259 \\ 
FCC & 12 & 0.908 & 1 & 0.816 & 1.113 & 1.225 \\
\br
\endTable

To conclude, in this letter we have developed a method of calculating the
correlation contribution to the free energy of a system which undergoes the
order-disorder transition. The essence of this method is the expansion in
powers of cumulant averages as applied to the exact configurational entropy.
It has been shown that this approach leads to the construction of simple 
polynomial approximations for the entropy. First corrections to the MFA free 
energy (see equation~(\ref{eq21}) for the correlation entropy) have been
calculated without any assumptions about the thermodynamic phase or the
character and range of the interactions in the system. Previously, the 
same idea was utilized by the author~\cite{Tsatskis-EPL} to simplify the 
videly used CVM by eliminating redundant variables in the CVM variational 
entropy. However, the approach presented here is more consistent, since 
the smallness of the pair and higher-order cumulants is the only assumption 
used, while in~\cite{Tsatskis-EPL} this assumption was combined with the 
very different CVM approximation. The advantage of both proposed methods 
is that the entropy polynomial needs to be derived once and for all for 
a given Hamiltonian and type of order; after that thermodynamic calculations 
become extremely easy and can be performed either analytically or using a 
pocket calculator, at least in relatively simple cases. For the quadratic 
approximation considered here this has been shown by the above example of 
the NN Ising ferromagnet. The SRO corrections lead to considerably better 
agreement, in comparison with the MFA, of the approximate transition 
temperatures with the corresponding best-known values. Calculation of 
the higher-order terms of the cumulant expansion for the configurational 
entropy is less straightforward, though perfectly possible, and is beyond 
the scope of the present short letter; this issue will be discussed in 
a separate publication~\cite{Tsatskis-to-be-published}.

\end{document}